\documentclass[10pt]{article}

\usepackage{epsfig}
\usepackage{subfigure}
\usepackage{amsmath}
\usepackage{amsthm}
\usepackage{amsfonts}
\usepackage{amssymb}

\title{A conjecture on the distribution of firm profit}
\author{Ian Wright\footnote{
iKuni Inc., 3400 Hillview Avenue, Building 5, Palo Alto, CA 94304, USA. Email: 
wrighti@acm.org, URL: ianusa.home.mindspring.com. Fax: +1 650 320 9827. Phone: +1 650 739 5355.}
\footnote{I am grateful to Julian Wells for explaining his work on firm profits, and
an anonymous reviewer for helpful criticisms.}}
\date{}

\begin{document}

\maketitle

\begin{abstract}
A common assumption of political economy is that profit rates across
firms or sectors tend to uniformity, and often models are formulated
in which this tendency is assumed to have been realised. But in
reality this tendency is never realised and the distribution of firm
profits is not degenerate but skewed to the right.  The mode is less
than the mean and super-profits are present. To understand the
distribution of firm profits a general probabilistic argument is
sketched that yields a candidate functional form. The overall
properties of the derived distribution are qualitatively consistent
with empirical measures, although there is more work to be done. \\ \\
Key words: firms, profit, economic, distribution, probabilistic
\end{abstract}

\section{Introduction}

Farjoun and Machover \cite{farjoun}, dissatisfied with the concept of
mechanical equilibrium applied to political economy and the
concomitant assumption of a realised uniform profit rate, outlined a
probabilistic approach to political economy, which replaced mechanical
equilibrium with statistical equilibrium and a uniform profit rate
with a distribution of profit rates. They reasoned that the proportion
of industrial capital, out of the total capital invested in the
economy, which finds itself in any given profit bracket will be
approximated by a gamma distribution, by analogy with the distribution
of kinetic energy in a gas at equilibrium. The gamma distribution is a
right-skewed distribution. They examined UK industry data from 1972
and concluded that it was consistent with a gamma distribution.  Wells
\cite{wells01} examined the distributions of profit rates defined in a
variety of ways of over 100,000 UK firms and found right-skewness to
be prevalent, but did not investigate their functional form. Wright
\cite{wright04a} measured the distribution of firm profits in an
agent-based model of a competitive economy, and found that the
distribution was right-skewed, although not well characterised by a
gamma distribution, even when capital-weighted. Analysis of the model
suggested that the profit distribution may be explained by general
probabilistic laws.

The remainder of the paper outlines some theoretical assumptions and
derives a candidate functional form for the distribution of firm
profits.

\section{A probabilistic argument}

Under normal circumstances a firm expects that a worker adds a value
to the product that is bound from below by the wage.  A firm's markup
on costs reflects this value expectation, which may or may not be
validated in the market. Wages are normally paid in installments of
between a week and one month, but the markup on costs is validated in
the market at a frequency that depends on the rate at which a firm's
goods and services are purchased by buyers. The frequency of payments
to a firm differ widely and depend on the complexity of the product
and the details of payment schedules (for example, compare a firm that
sells sweets to a firm that sells battleships).  The frequency
mismatch between wage payments and revenue payments can be mitigated
in many different ways, not least by the arrangement of capital
loans. But whatever the frequency of sale or the complexity of the
product a revenue payment to a firm partially reflects the value added
by the firm's workers during a period of time. Assume that the revenue
from the sale of a firm's product consists of a sum of market samples
where each sample represents the value-added by a particular employee
working for a small period of time, say an hour.  Obviously, there are
multiple and particular reasons why an individual worker adds more or
less value to the firm's total product, most of which are difficult to
measure, as partially reflected in the large variety of contested and
negotiable compensation schemes. Although each worker normally adds
value there is a great deal of local contingency. A worker may be a
slacker or a workaholic, an easily replaceable administrator, or a
unique, currently fashionable film star. Therefore, the precise value
contribution of an individual worker to the product is highly complex
and largely unknown, particularly when it is considered that the
productive co-operation of many workers cannot be easily reduced to
separate and orthogonal contributions, as is the case in highly
creative industries with production processes that have yet to mature
into separable, repeatable and well-defined tasks. This local
contingency and indeterminacy is modelled by assuming that the
value-added per worker-hour is a {\em random variable}. Consider that
a worker $i$ adds a monetary value, $X_{i}$, to a firm's product for
every hour worked, where each $X_{i}$ is an independent and
identically distributed (iid) random variable, with mean $\mu_{X}$ and
variance $\sigma_{X}^{2}$. The added value is assumed to be globally
idd to reflect the common determinants of the value-creating power of
an hour of work, but also random to model local
contingencies. Negative $X_{i}$ represents negative value-added,
corresponding to cases in which the worker's labour reduces the value
of inputs, for example the production of unwanted goods, or a slower
than average work pace, and so forth.

Assume that the distribution of $X_{i}$ is such that the Central Limit
Theorem (CLT) may be applied. Consider a single firm that sets in
motion a total of $n$ worker-hours during a single year. The firm's
total value-added, $S_{n}$, may therefore be approximated by a normal
distribution $S_{n} = \sum_{i=1}^{n} X_{i} \approx N( n \mu_{X}, n
\sigma_{X}^{2} )$.  The CLT approximation will improve with the size
of the firm, but even for small firms the number of iid draws is
large given the stated assumptions.

In reality the productivity of workers within firms is correlated. For
example, employees of firms that employ state-of-the-art machinery, or
are exceptionally well-organised, will all tend to add more value than
employees of firms that employ out-of-date machinery or are badly
organised.  Although competitive processes tend to homogenise the
value-added per worker, new innovations never cease, so that at any
moment in time the employees of particular firm will be more or less
productive than the average. A more accurate representation of
value-added is obtained if each $X_{i}$ is considered to be drawn from
a distribution indexed by the firm that employs worker $i$, at the
expense of a considerable increase in model complexity.  However, the
correlation of value-added within a large firm, which employs diverse
skills and machinery to produce a variety of products, will be weak.
Although a huge multinational is normally considered a single entity
for the purpose of reporting profits, in reality it sets into a motion
a large sample of different kinds of labours utilising different kinds
of machinery and tools. Hence, for large firms the assumption that
$X_{i}$ is sampled from a single, economy-wide distribution is a
reasonable approximation, for small firms less so. An advantage of
modelling value-added per worker as a random variable is that it is
possible that total value-added by a firm, $S_{n}$, is much higher or
lower than the norm, but this event has low probability. The
assumption of a single distribution that determines the value-added
per worker is able to approximate the diverse productivities of
individual firms.

Each worker costs a certain amount to employ during the year. This
cost includes the wage, the cost of inputs used by the worker, the
cost of wear and tear on any fixed capital, the cost of rent, local
taxes and so forth, all of which may be differently reported due to
local accountancy practices. Again, there is a great deal of
contingency. Hence costs per worker-hour are also modelled as a random
variable. Assume that a worker $i$ costs a monetary value, $Y_{i}$, to
productively employ per hour worked, where each $Y_{i}$ is an idd
random variable with mean $\mu_{Y}$ and variance $\sigma_{Y}^{2}$.
This cost includes both the wage and capital costs per worker, and
therefore effaces the distinction between variable and constant
capital. Costs per worker-hour are also correlated at the firm level:
the employees of different firms productively combine a greater or
lesser amount of capital. A more accurate representation of costs
would therefore consider the distribution of constant capital across
firms conditional on local circumstances, such as firm size, but this
extension is not pursued here. The assumption that cost per worker-hour is
statistically unifrom across firms is an approximation, which, as for
the case of value-added, improves with firm size, under the assumption
of a tendency toward homogenisation due to competitive pressures.

Assume that the distribution of $Y_{i}$ is such that the CLT may be
applied. Hence a firm that sets in motion $n$ worker-hours during a year has
total costs that may be approximated by a normal distribution, $K_{n}
= \sum_{i=1}^{n} Y_{i} \approx N( n \mu_{Y}, n \sigma_{Y}^{2} )$.
This approximation also improves with the size of the firm.

Different firms employee different numbers of workers and hence the
amount of hours worked for each firm during a year will vary.
Define the profit, $P_{n}$, of a firm that sets in motion $n$ hours of
labour in a single year as the ratio of value-added to costs,
$P_{n}=S_{n}/K_{n}$, and assume that $S_{n}$ and $K_{n}$ are
independent. $P_{n}$ is the ratio of two normal variates. Its
probability density function (pdf) may derived by the transformation
method (or alternatively see \cite{marsaglia65}) to give:

\begin{eqnarray}
\nonumber f_{P_{n}} (p \mid n) &=& 
\frac{\sqrt{n}  \exp[
-\frac{1}{2} n 
(\mu_{X}^{2} / \sigma_{X}^{2} + \mu_{Y}^{2} / \sigma_{Y}^{2})] }
{4 \pi (\sigma_{X}^{2} + p^{2} \sigma_{Y}^{2})^{3/2}} \\
& & \left(  \frac{2}{\sqrt{n}} \sqrt{\lambda_{1}} + \sqrt{2 \pi} \exp[\frac{n}{2} \frac{\lambda_{2}^{2}}{\lambda_{1}}] \lambda_{2}
\left( 1 + \Phi[\sqrt{\frac{n}{2}} \frac{\lambda_{2}}{\sqrt{\lambda_{1}}}] \right)
\right)
\label{eq:conditional}
\end{eqnarray}
where
\begin{eqnarray}
\nonumber \lambda_{1} & = & \sigma_{X}^{2} \sigma_{Y}^{2} ( \sigma_{X}^{2} + p^{2} \sigma_{Y}^{2} ) \\
\nonumber \lambda_{2} & = & \mu_{Y} \sigma_{X}^{2} + p \mu_{X} \sigma_{Y}^{2} \\
\nonumber \Phi(x) & = & \frac{2}{\sqrt{\pi}} \int_{0}^{x} \exp^{-t^{2}} dt
\end{eqnarray}

Equation (\ref{eq:conditional}) is the pdf of the rate-of-profit of a
firm conditional on $n$, the number of hours worked for the firm per
year.

Axtell \cite{axtell} analysed US Census Bureau data for US firms
trading between 1988 and 1997 and found that the firm size
distribution, where size is measured by the number of employees,
followed a special case of a power-law known as Zipf's law, and this
relationship persisted from year to year despite the continual birth
and demise of firms and other major economic changes. During this
period the number of reported firms increased from 4.9 million to 5.5
million.  Gaffeo et. al. \cite{gaffeo03} found that the size
distribution of firms in the G7 group over the period 1987-2000 also
followed a power-law, but only in limited cases was the power-law
actually Zipf.  Fuijiwara et. al. \cite{fujiwara03} found that the
Zipf law characterised the size distribution of about 260,000 large
firms from 45 European countries during the years 1992--2001. A Zipf
law implies that a majority of small firms coexist with a decreasing
number of disproportionately large firms. Firm sizes theoretically
range from 1 (a degenerate case of a self-employed worker) to the
whole available workforce, representing a highly unlikely
monopolisation of the whole economy by a single firm.

The empirical evidence implies that at any point in time the firm size
distribution follows a power-law, and that this distribution is
constant, despite the continual churning of firms in the economy
(birth, death, shrinkage and growth). Firms hire and fire employees,
and therefore the number of hours worked for a firm during a year
depends on its particular historical growth pattern. To simplify,
assume that the average number of employees per firm per year also
follows a power-law. This approximation is reasonable if the growth
trajectories of firms do not fluctuate too widely during the
accounting period. Assume also that every employee works the same
number of hours in a year, which is a reasonable simplification. The
firm hours per year is therefore a constant multiple of the number of
firm employees. Firms with more employees proportionately set in
motion more hours of labour.  A constant multiple of a power-law
variate is also a power-law variate. Hence, the firm size distribution
has the same power-law form whether firm size is measured by employees
or by the total number of hours worked by employees.  

The unconditional rate-of-profit distribution can therefore be
obtained by considering that the number of hours worked for a firm
during a year is a random variable $N$ distributed according to a
Pareto (power-law) distribution:
\begin{eqnarray}
\nonumber f_{N}(n) = \frac{\alpha \beta^{\alpha}}{n^{\alpha + 1}}
\end{eqnarray}
where $\alpha$ is the shape and $\beta$ the location parameter. Assume
that firm sizes range between $m_{1}$ hours, which represents a
degenerate case of a self-employed worker who trades during the year,
to $m_{2}$ hours, which represents a highly unlikely monopolisation of
all social labour by a single huge firm ($m_{2} >> m_{1}$). The
truncated Pareto distribution
\begin{eqnarray}
\nonumber g_{N}(n) = f_{N}( n \mid m_{1} < N \leq m_{2} ) = 
\frac{f_{N}(n)}{F_{N}(m_{2}) - F_{N}(m_{1})} = 
\frac{n^{-(1 + \alpha)} \alpha m_{1}^{\alpha} m_{2}^{\alpha}}
{m_{2}^{\alpha} - m_{1}^{\alpha}}
\end{eqnarray}
where 
\begin{eqnarray}
\nonumber f(n) = F'(n)
\end{eqnarray}
is formed to ensure that all the probability mass is between $m_{1}$
and $m_{2}$.  Assume that $m_{2}$ is large so that the discrete firm
size distribution can be approximated by the continuous distribution
$g_{N}$.

By the Theorem of Total Probability the unconditional profit distribution
$f_{P}(p)$ is given by:
\begin{equation}
f_{P}(p) = \int_{m_{1}}^{m_{2}} f_{P}(p \mid n) g_{N}(n) dn
\label{eq:unconditional}
\end{equation}
Expression (\ref{eq:unconditional}) defines the $g_{N}(n)$
parameter-mix of $f_{P}(p \mid N = n)$.  The rate-of-profit variate is
therefore composed of a parameter-mix of a ratio of independent normal
variates each conditional on a firm size $n$, measured in hours per
year, distributed according to a power-law. Writing
(\ref{eq:unconditional}) in full yields the pdf of firm profit:

\begin{eqnarray}
\nonumber f_{P} (p) &=& \int_{m_{1}}^{m_{2}}
\frac{\exp[
-\frac{1}{2} n 
(\mu_{X}^{2} / \sigma_{X}^{2} + \mu_{Y}^{2} / \sigma_{Y}^{2})] }
{4 \pi (\sigma_{X}^{2} + p^{2} \sigma_{Y}^{2})^{3/2}} \\
\nonumber& & \left(  \frac{2}{\sqrt{n}} \sqrt{\lambda_{1}} + \sqrt{2 \pi} \exp[\frac{n}{2} \frac{\lambda_{2}^{2}}{\lambda_{1}}] \lambda_{2}
\left( 1 + \Phi[\sqrt{\frac{n}{2}} \frac{\lambda_{2}}{\sqrt{\lambda_{1}}}] \right)
\right) \\
& & \frac{n^{-(\frac{1}{2} + \alpha)} \alpha m_{1}^{\alpha} m_{2}^{\alpha}}
{m_{2}^{\alpha} - m_{1}^{\alpha}} \; dn
\label{eq:unconditionalFull}
\end{eqnarray}

This distribution has 7 parameters: (i) $\mu_{X}$, the mean
value-added per worker-hour, (ii) $\sigma_{X}^{2}$, the variance of
value-added per worker-hour, (iii) $\mu_{Y}$, the mean cost per
worker-hour, (iv) $\sigma_{Y}^{2}$, the variance of cost per
worker-hour, (v) $\alpha$, the Pareto exponent of the firm size
power-law distribution, where size is measured in worker-hours per
year, (vi) $m_{1}$, the number of hours worked by a single worker in a
year, and (vii) $m_{2}$, the total number of hours worked in the whole
economy during a year. Both percentage profit, $R = 100 P$, and the
growth rate of capital invested, $G = 1 + P$, are simple linear
transforms of this distribution.

The parameters can be estimated from economic data and the resulting
distribution compared to empirical rate-of-profit measures, under
various simplifying assumptions about how profit is defined (e.g.  see
Wells \cite{wells01}). A good fit would imply that the assumptions
made in the theoretical derivation are empirically
sound. Alternatively, best-fit parameters may be directly estimated
from empirical data, for example by the method of maximum likelihood
estimation, to determine how well the theoretical distribution can fit
a set of empirical distributions. A good fit compared to other
candidate functional forms would imply that a parameter-mix of a ratio
of normal variates with parameters conditional on a power-law captures
some essential structure of the determinants of firm profit, but it
would not validate the theoretical derivation.

\begin{figure}
\centering \nonumber \subfigure[{Samples plotted on separate
scales.}]{\epsfig{file=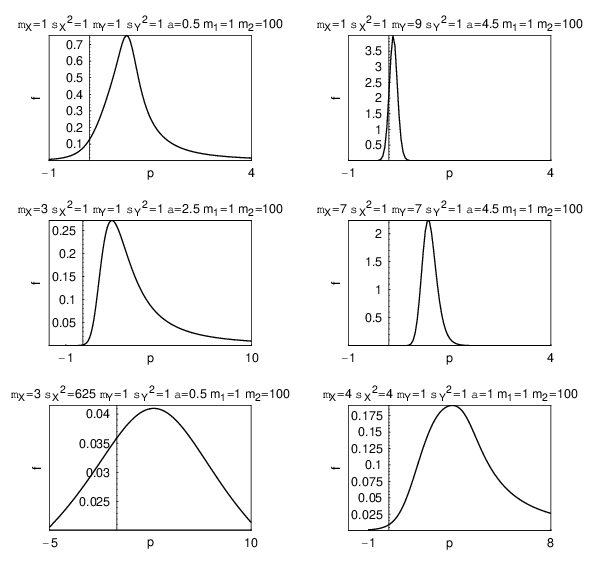,viewport=0 0 300
270,width=0.9\textwidth,clip=true,silent=}}


\subfigure[{The same samples plotted on a single scale.}]{\epsfig{file=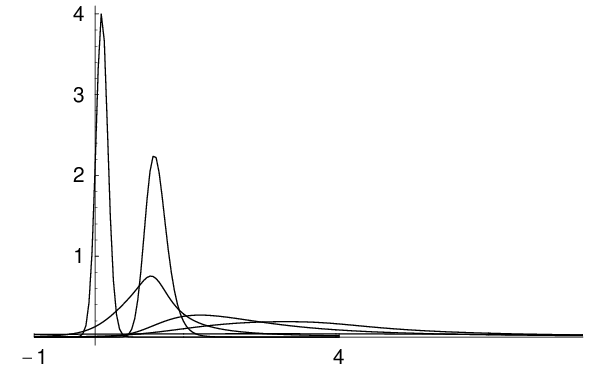,viewport=0 0 260 180,width=0.5\textwidth,clip=true,silent=}}\\

\caption{Representative numerical samples of the probability density
function $f_{P}(p)$.}
\label{fig:samples}
\end{figure}

Equation (\ref{eq:unconditionalFull}) is difficult to analyse so numerical
solutions are employed. Figure 1 graphs some representative numerical samples of the 
distribution. The samples range from sharply peaked symmetrical curves, 
in which most of the probability mass is concentrated about the mode, 
to less peaked distributions that are skewed to the right. Wells' \cite{wells01}
variety of profit measures yield distributions that share these characteristics,
and therefore there is qualitative agreement between the theory and the
empirical data. But clearly a full quantitative analysis is required.

\begin{figure}[h]
\begin{center}
\epsfig{file=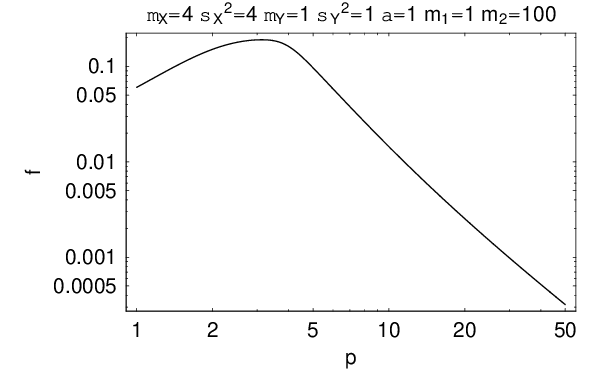,viewport=0 0 280 200,width=.7\textwidth,clip=true,silent=}
\caption{A sample of $f_{P}(p)$ plotted in log-log scale. Note the long power-law tail.}
\end{center}
\label{fig:longTail}
\end{figure}

Figure 2 graphs a sample of $f_{P}(p)$ in log-log scale. The
approximate straight line in the tail is the signature of a power-law
decay of the probability of super-profits.  Super-profit outliers are
found in the empirical data, although it has not been investigated
whether they decay as an approximate power-law.

Further analysis of the pdf $f_{P}(x)$ is required. But the
qualitative form of the distribution is sufficiently encouraging to
consider it a candidate for fitting to empirical profit measures and
for comparison with other candidate functional forms. To go beyond
models that assume a realised uniform profit rate it is necessary to
investigate empirical data on firm profit and propose theoretical
explanations of its distribution.  This paper is a tentative step in
that direction.

\section{Conclusion}

A general probabilistic argument suggests that the empirical 
rate-of-profit distribution will be consistent with a parameter-mix 
of a ratio of normal variates with means and variances that depend 
on a firm size parameter that is distributed according to a power law. 

\bibliography{ian}
\bibliographystyle{plain}

\end{document}